# Development of a Scalable Quantum Memory Platform—Materials Science of Erbium-Doped TiO$_2$ Thin Films on Silicon

Manish Kumar Singh[1], Gary Wolfowicz[2], Jianguo Wen[3], Sean E. Sullivan[2], Abhinav Prakash[3], Alan M. Dibos[2,3], David D. Awschalom[1,2], F. Joseph Heremans[1,2], Supratik Guha[1,2]*

[1]Pritzker School of Molecular Engineering, University of Chicago, Chicago, IL, USA 60637

[2]Materials Science Division, Argonne National Laboratory, Lemont, IL, USA, 60439

[3]Center for Nanoscale Materials, Argonne National Laboratory, Lemont, IL, USA, 60439

*Corresponding author: Supratik Guha (guha@uchicago.edu, sguha@anl.gov)

## Abstract

Rare-earth ions (REI) have emerged as an attractive candidate for solid-state qubits, particularly as a quantum memory. Their 4f-4f transitions are shielded by filled 5s and 5p orbitals, offering a degree of protection from external electric fields. Embedded within a thin film oxide host, REIs could enable a qubit platform with significant memory capabilities. Furthermore, a silicon-compatible thin film form factor would enable the use of standard semiconductor fabrication processes to achieve chip-based integrability and scalability for functional quantum networks. Towards this goal, we have carried out optical and microstructural studies of erbium-doped polycrystalline and epitaxial $TiO_2$ thin films on Si(100), r-sapphire, and $SrTiO_3$(100). We observe that the inhomogeneous optical linewidth of the Er photoluminescence is comparable or better for polycrystalline Er: $TiO_2$ (grown on Si) in comparison to single crystal epitaxial films on sapphire or $SrTiO_3$, implying a relative insensitivity to extended defects. We investigated the effect of the film/substrate and film/air interface and found that the inhomogeneous linewidth and spectral diffusion can be significantly improved via bottom buffer and top capping layers of undoped $TiO_2$. Using such approaches, we obtain inhomogeneous linewidths of 5.2 GHz and spectral diffusion of 180 MHz in Er:$TiO_2$/Si(100) films and have demonstrated the engineerability of quantum-relevant properties in these thin films.

Quantum memories are a key part of emerging quantum communication networks that establish entanglement-based links over long distances[1–6]. They are also anticipated to be an essential component for quantum processors that require the storage of quantum states while carrying out a system task. Solid-state rare-earth memories are of interest because the form factor offers ease of handling and the potential for massive scalability, and rare-earth ions (REIs) are competitive candidates for quantum memory. In particular, embedding the REI qubits in thin-film wide band-gap solid-state hosts has emerged as a promising platform[3,7,8], due to their compatibility with established semiconductor industry processes and potential for on-chip integration. One important driving force in this area has been demonstrations of interaction with single atoms using nanophotonic structures for Purcell enhanced increase in emission rates and optical access of the active rare earth ion[9,10].

Motivated by this, we present the growth and characterization of Er-doped polycrystalline $TiO_2$ thin films on silicon substrates as a promising scalable platform. As a comparison, we have also examined Er:$TiO_2$ films epitaxially grown on other substrates. The choice of $Er^{3+}$, whose first optical excited state emits in the telecom C-band (1530 - 1565 nm), has been described by us elsewhere[11]. The choice of $TiO_2$ thin films is motivated by its compatibility with Si process technology, wide band-gap (3.2 eV), and an inert near-nuclear spin-free (87%) host environment for Er. $TiO_2$ is also relatively free of trace rare earth impurities. Recent investigations [12] have revealed promising optical properties of erbium emission in bulk $TiO_2$.

We report insights into the host material's effect on the optical properties of erbium dopants. First, the polycrystalline $TiO_2$ films on silicon are shown to have narrower inhomogeneous linewidth compared to the single crystal films on lattice-matched substrates hinting at other factors influencing these properties. Second, the phase (rutile, anatase) can be controlled using growth temperature when growing on silicon. Third, we can engineer the inhomogeneous linewidth ($\Gamma_{inh}$) by using a film stack consisting of undoped $TiO_2$ bottom buffer and top capping layers. Finally, we show that the spectral diffusion linewidth can also be engineered using erbium concentration and the film stack.

# Methods
## A. Growth
Films of $TiO_2$ were vacuum-deposited using a Riber molecular beam epitaxy (MBE) system. Titanium tetraisopropoxide (TTIP) from Sigma-Aldrich with a purity of 99.999% (trace metal basis) was used as the titanium precursor; details of TTIP-based MBE of $TiO_2$ have been described in detail by Jalan *et al.*[13]. Erbium doping (10-1000 ppm range) was carried out by using a high-temperature effusion cell. Films were grown at substrate temperatures between 480-850 °C, molecular oxygen flow of 0.55 sccm, and growth rates of ~60 nm/hr. Substrates used were r-plane sapphire, $Al_2O_3$, $SrTiO_3$ (100) (STO), and Si (100). The STO and sapphire substrates were coated with 350 nm Ta on the backside for improved radiative heating. Prior to growth, the oxide substrates were cleaned in-situ using 20 min of $O_2$ plasma (325 W). For the Si substrates, we performed an RCA clean followed by an HF dip. The default doping concentration for the films discussed is estimated at ~35 ppm (unless otherwise specified) – concentration was estimated using a film grown using an erbium cell temperature 1200 °C and extrapolating the vapor pressure[14,15] to the cell temperature used for doping.

## B. Characterization
The growth and substrate reconstruction were monitored in-situ using reflection high-energy electron diffraction (RHEED). Ex-situ film characterization was performed using X-ray diffraction (XRD), Atomic

Force Microscopy (AFM), transmission electron microscopy (TEM), and optical spectroscopy. Electron microscopy characterization used the Argonne Chromatic Aberration-corrected TEM (ACAT, an FEI Titan 80-300ST) and an FEI Talos 200X microscope.

Optical spectroscopy measurements were performed in a custom-built cryogenic confocal microscope using a closed-cycle cryostat with a base temperature of 3.5K (unless otherwise noted). Laser light was focused on the sample using an external 50x objective (NA 0.67). Off-resonant measurements were performed using a 905 nm excitation laser (QPhotonics QFLD-905-200S). Broadband $Er^{3+}$ photoluminescence was collected with a fiber-coupled spectrometer and a liquid nitrogen-cooled InGaAs camera (IsoPlane-320 and PyLoN-IR, Teledyne Princeton Instruments) using a 300 g/mm diffraction grating blazed for 1.2 μm light, achieving a spectral resolution of approximately 0.3 nm. On resonant, photoluminescence excitation (PLE) spectroscopy was performed using a tunable C-band laser (PurePhotonics). Light was collected (emitted) from (into) free-space from a single collimator into a polarization-maintaining fiber. The fiber is connected to a bidirectional 2x1 acousto-optic modulator switch (Brimrose) which switches between a collection and an excitation path. During excitation, the switch routes laser light onto the sample. During collection, the switch routes the photoluminescence to a fiber-coupled superconducting nanowire single-photon detector (SNSPD, Quantum Opus). Two additional acousto-optic modulators (AA-Optoelectronic) on the excitation path fully attenuate any cross-talk from the excitation laser to well below the SNSPD dark noise (< 100 counts/s). Spectra are obtained by sweeping the wavelength of the tunable laser and exciting for a pulse duration between 0.1 and 1 ms. The photoluminescence is collected after a buffer delay of 100 μs after the excitation pulse and is measured for a duration between 1 and 15 ms. Spectral diffusion measurements are realized using an intensity electro-optic modulator (10 GHz, Thorlabs), biased at the quadratic point, to create the probe sidebands. The sideband frequency is set and swept by a driving external microwave signal generator.

## Results and Discussion

### A. Film Structure

$TiO_2$ crystallizes into two main phases, anatase ($I4_1/amd$, a = 3.7845 Å, c = 9.5143 Å; Z = 4) and rutile ($P4_2/mnm$ a = 4.5937 Å, c = 2.9587 Å; Z = 2). Of these, rutile has been grown epitaxially on r-plane sapphire with a mismatch of 3.7% along $[010]_{TiO_2}$ and 6.04% along $[\bar{1}01]_{TiO_2}$[16]. Similarly, STO substrates have been used for the stabilization of the anatase phase with a lattice mismatch of 3.1%[17].

Our results for both $TiO_2$ phases grown on r-plane sapphire, STO (100), and Si (100) are summarized in the cross-sectional TEM results of Figure 1. Streaky RHEED images observed during deposition (see supplementary information, SI) and selected area electron diffraction (SAED) patterns from the TEM analysis (figs 1(i), (k)) shows that $TiO_2$ films deposited on sapphire and STO substrates are single crystal. The SAED further confirms a rutile structure for the $TiO_2$ growth on r-sapphire and a crystallographic orientation relationship of: $[101]_R//[211]_S, (\bar{1}01)_R//(10\bar{2})_S, (020)_R//(1\bar{2}0)_S$ with subscripts S and R indicating sapphire and rutile, respectively. On the STO substrate, we observe anatase films growing epitaxially with the crystallographic orientation relationship: $[100]_A//[001]_S, [010]_A//[100]_S, [001]_A//[010]_S$ – with subscripts A and S indicating anatase and STO. Microstructural observations indicate epitaxial growth and the absence of a discernible interfacial layer between the film and substrate for both systems.

Thin films grown on silicon substrates at 750 °C are polycrystalline rutile with a grain size of ~50 nm (Figure 1(b)). The film/substrate interface is wavey (see the HRTEM image of Figure 1(f)), and a 3-4 nm thick interfacial layer is observed and determined to be $SiO_x$ by X-ray energy dispersive spectroscopy (EDS) (see SI). Integrated SAED patterns taken from 25 different areas on the film show that there is a degree of epitaxial relationship of rutile $TiO_2$ on Si (further details are provided in the SI). This suggests that the interfacial $SiO_x$ is formed following the growth of the film, likely via diffusion of oxygen through the $TiO_2$ layer, as is known in other dielectric oxides[18].

When the deposition temperature is lowered to 480 °C, the film grown on Si is found to be polycrystalline anatase with columnar grains of size ~10 nm (Figure 1(d)). HRTEM images show that the interfacial layer between film and substrate is less than 1 nm. Integrated SEAD patterns from 25 areas on the film (SI) suggests a certain degree of epitaxial relationship of anatase $TiO_2$ on Si.

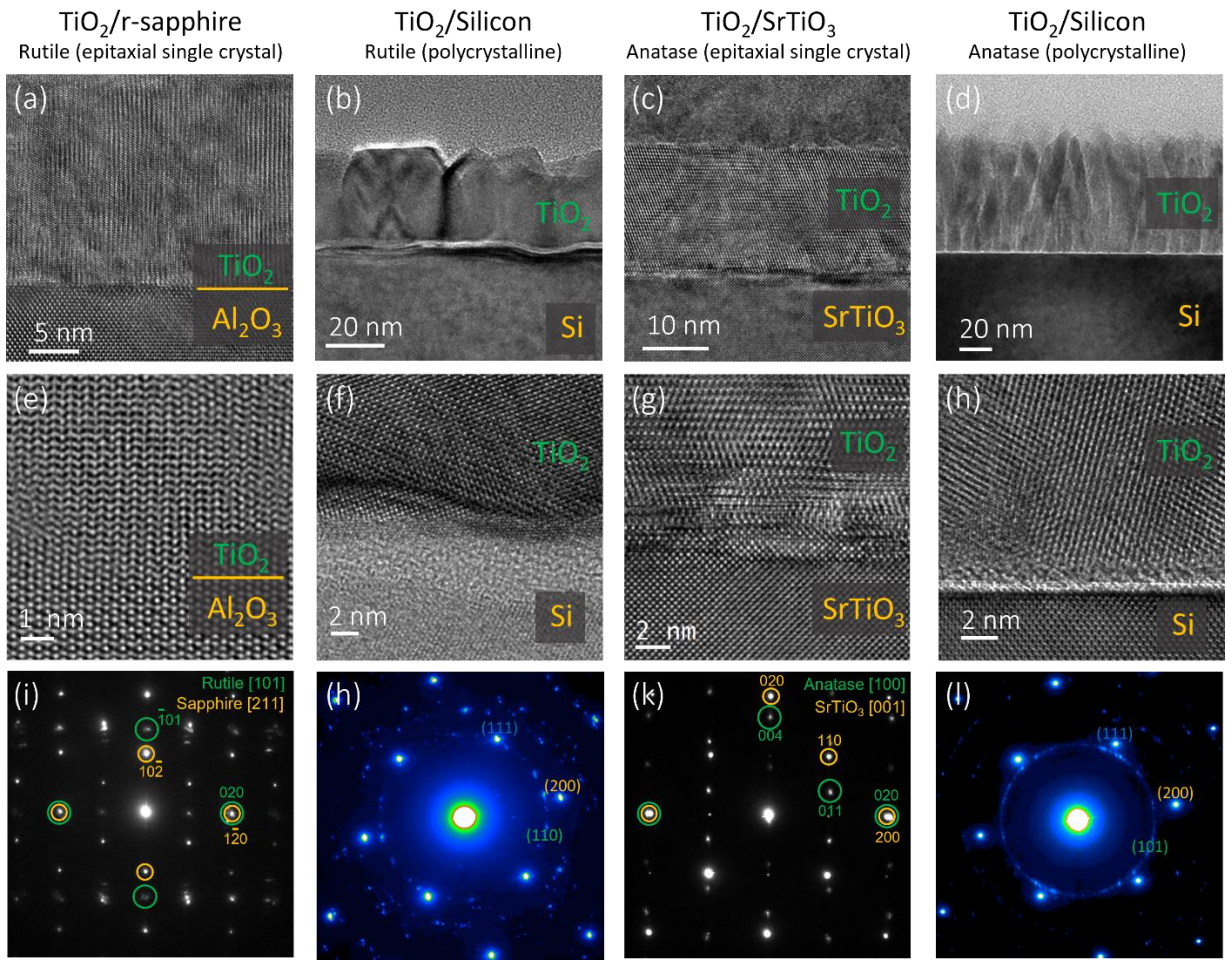

*Figure 1. TEM data for the $TiO_2$ films/substrates investigated are shown here – growth on STO and sapphire shows an epitaxial relationship and single crystal growth while growth on Si is polycrystalline. (a)-(d) TEM showing the films on the substrates (as labeled) (e)-(h) magnified image of the interface shows how the interface compares in the four cases, showing epitaxy (e & g) and showing an oxide interface (f & h). (i-l) Selected area electron diffraction (SAED) confirms the epitaxial relationship for sapphire (i) and STO (k). For growth on silicon, we find the film is textured along a preferred growth direction, as also evidenced from the observed XRD peaks (SI).*

## B. Spectroscopic characterization

For single crystal rutile (on sapphire), we obtain a prominent erbium emission peak ~1520 nm corresponding to the $^4I_{13/2}$ → $^4I_{15/2}$ transition, consistent with earlier observations in Er:TiO$_2$ (bulk rutile)[12] at low concentrations. For single-crystal anatase (on STO), the prominent erbium emission peak is at ~1533 nm, in agreement with prior literature[19,20] ($^4I_{13/2}$ → $^4I_{15/2}$ transition). As shown in Figure 2, emission peaks in the same region are seen for films grown on silicon for polycrystalline rutile (high growth T) and anatase (low growth T) – also confirmed by TEM in Figure 1. Intermediate growth temperatures show peaks corresponding to both phases.

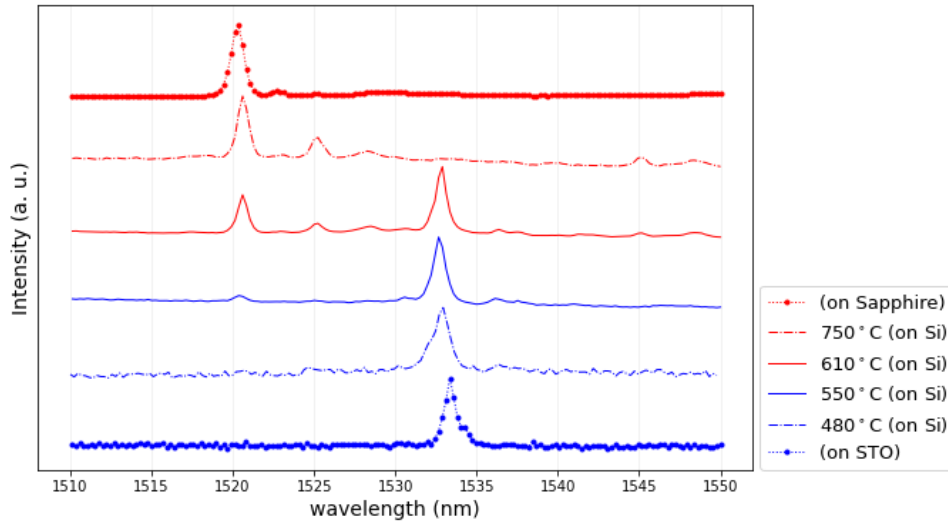

*Figure 2. Photoluminescence (PL) spectra of Er:TiO$_2$ on silicon grown at different temperatures. Er$^{3+}$ in rutile TiO$_2$ has an emission wavelength of ~1520 nm and in anatase of ~1533 nm. As the growth temperature increases from 480 °C to 750 °C, the rutile peak intensity increases while the anatase peak decreases. PL data from Er:TiO$_2$ on sapphire and STO is included for reference. Note that the offset is artificial, added for clarity.*

In Figure 3, we compare the inhomogeneous photoluminescence (PL) spectra and resulting inhomogeneous linewidths, $\Gamma_{inh}$, (using Lorentzian fits) for four types of TiO$_2$ films: single crystal and polycrystalline TiO$_2$ in the rutile phase grown on r-sapphire and Si(100) substrates respectively (Figure 3a); and single crystal and polycrystalline TiO$_2$ in the anatase phase grown on STO and Si(100) substrates respectively (Figure 3b). Figure 3(a) shows the inhomogeneous linewidth of the rutile films to be 109±5 GHz (single crystal on sapphire) and 91±5 GHz (polycrystalline on Si). Both samples were through doped (i.e., no undoped buffer layer on top or bottom), the film thickness was ~30nm, and the estimated erbium concentration was 35 ppm. As a comparison, the inhomogeneous linewidths of erbium-doped (via implantation) bulk rutile has been reported as 460 MHz[12]. The anatase films have inhomogeneous linewidths of 18.9±0.8 GHz (single crystal on STO) and 11.1±0.2 GHz (polycrystalline on Si). Both anatase films have a buffered structure (10 nm undoped bottom buffer/10 nm doped layer/5 nm undoped cap) with a dopant concentration of 35 ppm (discussed later).

There are two key observations to be made. Firstly, the anatase films have a narrower inhomogeneous optical linewidth than the rutile films. We speculate that differences in local symmetries and point defects such as bulk and surface/near-surface vacancies may play a role. Oxygen vacancies (which can be charged) are known to be a dominant point defect in a wide variety of wide gap metal oxides, including TiO$_2$[21,22].

Theoretical studies have suggested, for instance, that the energy barrier for vacancy formation is higher for anatase (101) compared to rutile (110) surface[23]. In experiments, 5-10% concentration of vacancies is obtained for the rutile (110) surface[24] while it is reported to be <1% for the anatase (101) surface[22,25]. We note that the use of undoped buffer and cap layers for the anatase films can possibly explain some of the improvement in linewidth observed. However, as will be indicative from results discussed later in this paper (as well as our prior work[11] on doping sequence variations in $Y_2O_3$ films), this cannot likely account for the ~10X difference seen between the rutile and anatase $TiO_2$ films.

Our second observation – the lack of linewidth improvement going from single to polycrystalline films – indicates that extended defects such as grain boundaries are not playing a significant role on the PL linewidths. This conclusion is also supported by noting that the rutile films on Si have a larger grain size compared to the anatase films (from the TEM analysis) but a broader $\Gamma_{inh}$. These results would indicate that strain fields (and resulting electric fields) caused by grain boundaries/extended defects in the host oxide do not significantly impact the inner 4f excitations of the Er dopant, possibly due to the shielding by the outer orbitals. The nuclear spin environment and the dopant densities (~35 ppm) are the same in both cases—hence nuclear spin and dopant dipole-dipole interactions cannot explain the differences. On the other hand, we note that almost all wide band-gap oxides (including $TiO_2$) that are strongly ionic have a rich defect chemistry that arises from non-stoichiometry[21,22,26,27], leading to charged vacancies on the cationic or anionic sites and their complexes. The presence of non-uniform distributions of such charged defects in the vicinity of the erbium can lead to PL broadening. We speculate that a significant contribution to the broadening may be caused by the presence of such point defects. These fluctuations in the short-range environment mask the longer-range effects from extended defects.

The 20-50% improvements in going to polycrystalline films compared to single-crystal films is not well understood, though consistent with our above speculation. The presence of grain boundaries can be an effective getter for point defects by acting as a sink[28]. As a result, the immediate short-range environment of the Er remains "cleaner" due to this getter effect of the grain boundaries and dislocations, which sweeps these point defects away from the locality of the Er atoms, leading to a "cleaner" short-range environment.

In the following, we focus on polycrystalline anatase on silicon to study the effect of the interface, concentration, and sample temperature, as it offers the overall narrowest linewidth. Unless otherwise noted, the experiments results reported are for the anatase phase.

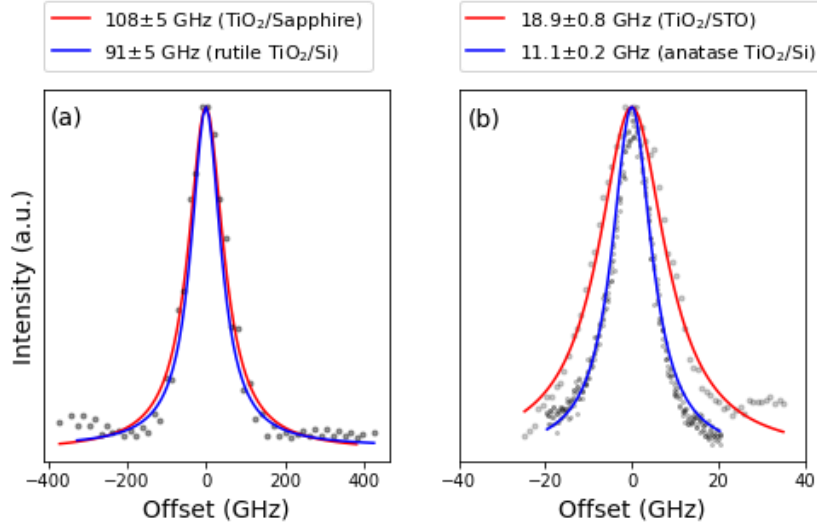

*Figure 3. (a) The rutile phase in both single crystal and polycrystalline forms shows inhomogeneous linewidths >90 GHz – the two films are otherwise identical (~30nm thickness, through doped, 35 ppm of erbium). (b) The anatase phase, by contrast, shows an inhomogeneous linewidth much smaller compared to that in the rutile phase. The polycrystalline anatase on silicon demonstrates a narrower linewidth compared to single-crystal anatase on STO (by ~ 41%). These two films are identical in other ways (10 nm bottom buffer/10 nm doped layer/5 nm top cap), 35 ppm erbium).*

## C. Engineering thin film properties

In our earlier investigations of epitaxial Er:$Y_2O_3$ on Si[11], we have shown that the distance of the erbium-doped layer from the substrate-film interface had a significant effect on $\Gamma_{inh}$. To explore this effect in $TiO_2$/Si(100), we carried out experiments using a film stack of undoped $TiO_2$ bottom buffer(x nm; x=10-60 nm)/Er:$TiO_2$ (10 nm)/undoped $TiO_2$ cap (5 nm). We use a 10nm doped thin film with no buffer or cap as reference. The results, plotted in Figure 4, show a reduction in $\Gamma_{inh}$ from ~31 GHz (for a 10nm film with no buffer or cap) to ~5.2 GHz with x= 60 nm and a capping layer of 5 nm. Taken together with our previous Er:$Y_2O_3$ results[11], this implies a broader underlying phenomenon that seems independent of whether the films are epitaxial or polycrystalline. Additionally, it might also apply to other systems, as adding a capping/buffer layer to the doped films has also been suggested by Harada *et al.*[29] for Eu doped $Y_2O_3$ films grown using chemical vapor deposition. These results would be consistent with higher concentrations of charged defects in the vicinity of the bottom substrate interface and film surface. Ionic metal oxide dielectrics are surface sensitive, and near-surface volumes are permeable to moisture[30] and, therefore, the presence of $OH^-$ ions. Electrochemically driven oxygen vacancy exchange (leading to interfacial silicon oxide formation) at the silicon-metal oxide interface[31] may lead to pileups in charged oxygen vacancies at the bottom interface.

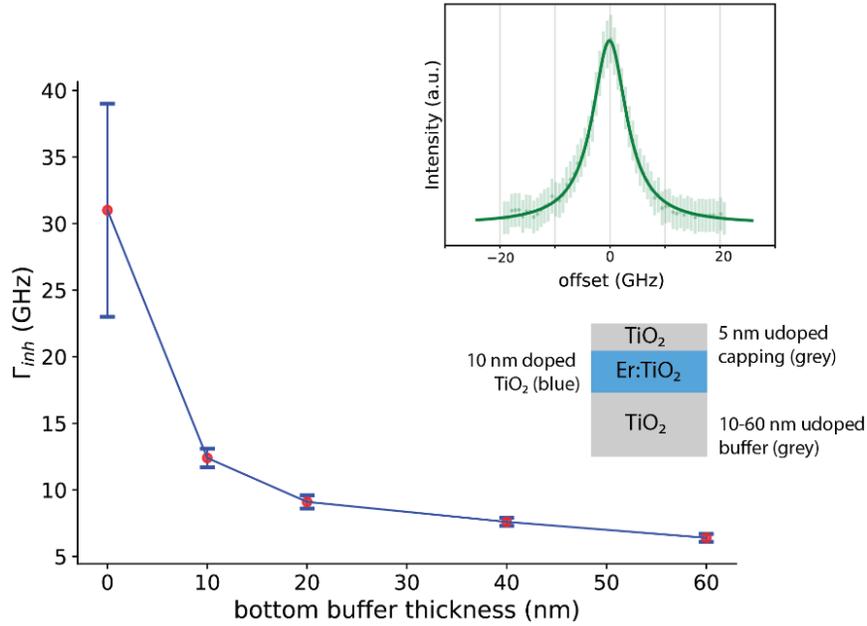

Figure 4. The inhomogeneous linewidth of Er:TiO$_2$ on silicon as a function of bottom buffer layer thickness between the silicon and the TiO$_2$. The narrowest linewidth was obtained for the 60-nm bottom buffer (inset). The schematic shows an erbium-doped thin film (blue) with a top cap and bottom buffer of undoped TiO$_2$ (grey). Note that the sample with no bottom buffer also has no top capping layer.

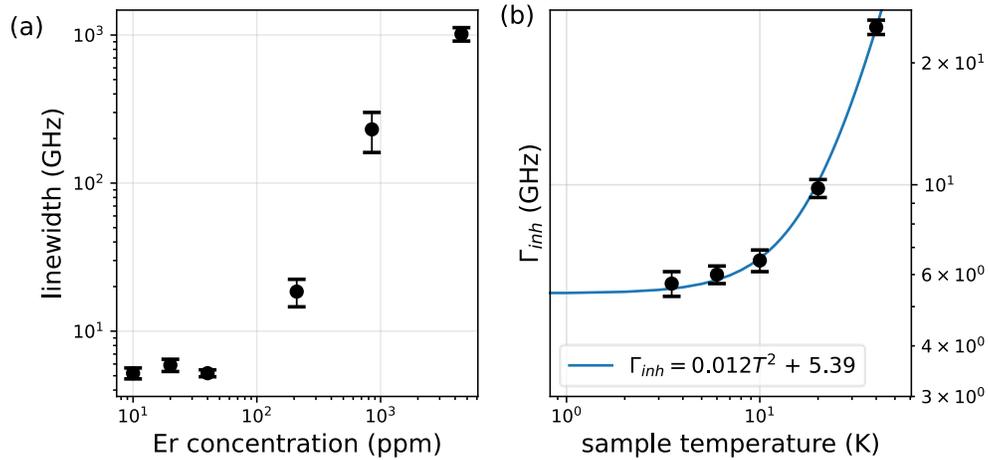

Figure 5 (a) Erbium concentration doesn't appear to be the limiting contributor to $\Gamma_{inh}$ We used (TiO$_2$ (40 nm, buffer)/ Er: TiO$_2$ (10 nm)/ TiO$_2$ (5 nm, capping)) stack for this set of experiments. (b) The contribution from the sample temperature also appears to be small at 3.5K – note the quadratic fit for this sample projects a minimum of 5.39 GHz. These experiments were performed using a (TiO$_2$ (40 nm, buffer)/ Er: TiO$_2$ (20 nm)/ TiO$_2$ (10 nm, capping)) stack.

We also explored the dependence of $\Gamma_{inh}$ on the erbium concentration in the films and the measurement temperature (during optical studies). For the erbium concentration dependence, we prepared a series of Si(100)/TiO$_2$ (40 nm)/ Er: TiO$_2$ (10 nm)/ TiO$_2$ (5 nm) samples with different erbium doping concentrations and with the substrate temperature at 480 °C (resulting in the anatase phase). The results of the measured $\Gamma_{inh}$ (Figure 5(a)) indicate that below ~100 ppm (ion-ion separation of ~4.8 nm), the inhomogeneous linewidth is not significantly degraded by concentration effects.

Figure 5(b) shows the temperature dependence (3.5K - 40 K) of the optical inhomogeneous linewidth for a Si(100)/TiO$_2$ (40 nm)/ Er: TiO$_2$ (20 nm)/ TiO$_2$ (10 nm) film. Note that at temperatures below 3.5 K, the linewidth narrowing plateaus off. The linewidth reduces with temperature and exhibits a quadratic temperature dependence given by: $\Gamma_{inh} = 0.012T^2 + 5.39$ where T is in kelvin and $\Gamma_{inh}$ in GHz. One consideration for this dependence is the known increase in the dielectric permittivity of TiO$_2$ with decreasing temperature[32], which has been modeled by Samara and Peercy for the rutile phase and evaluated more recently by Bonkerud *et al*.[33] The resultant increased dielectric screening of electric fields (from defects etc.) by the host media can then reduce the PL broadening of the ensemble of embedded REIs. However, this expected dependence does not fit the observed quadratic dependence, pointing to other factors playing a role, such as the quenching of electrically active states in the host with decreasing temperature.

We now turn to a measurement of the coherence times. While we were unable to perform the direct measurement of T2* (Ramsey experiment) or T2 (Hahn-echo) measurements due to the fast-dephasing times, we were able to extract an upper limit on T2* via an indirect method (described in the methods section) based on transient hole burning as described by Weiss *et al*.[34] We measured the spectral diffusion on several samples with varying bottom buffer thickness, capping layer thickness, and dopant concentration and the results are summarized in Figure 6. Three observations can be made regarding the spectral diffusion linewidth. Firstly, the spectral diffusion varies from 180 to 425 MHz. This is slightly larger than the spectral diffusion (45MHz and 110 MHz) reported for individual erbium atoms implanted in silicon[34]. There is an improvement with the buffer layer thickness, consistent with the results for the inhomogeneous PL linewidths. Secondly, there appears to be a small decrease in the spectral diffusion linewidth (~ ~20%) with a lower doping density (from 35 to ~10 ppm). Thirdly, an increase in the capping thickness from 10 to 20 nm yielded a large reduction in spectral diffusion linewidth (~50%). This is likely due to the higher sensitivity of spectral diffusion to charge defects and the proximity to the surface, consistent with our earlier arguments regarding the inhomogeneous linewidth variation. $NV^-$ centers in diamond are known to be sensitive to the depth from the surface and charge fluctuations[35]. We further note that such ionic oxides can be affected by moisture penetration from the ambient air[30].

## Conclusion

Several conclusions can be made from our examination of Er: TiO$_2$ thin films as a silicon-compatible host for solid-state quantum devices. Our study examined the growth, microstructure, and optical properties of these films under a variety of conditions, primarily on silicon substrates (but also on sapphire and STO substrates for comparison). Firstly, the PL linewidths seem fairly insensitive to the crystallinity (single vs. poly) of the host films and the presence of extended defects. It implies that PL quality and the optical linewidth are currently limited by other materials effects such as, possibly, charged point defects arising from non-stoichiometry. Combined with our previous results on Y$_2$O$_3$ films on silicon and literature, it may generally be applicable to other REI doped oxide thin film systems. Secondly, we show that the growth of a suitable buffer and a top capping layer is necessary for improved optical properties (both homogeneous and inhomogeneous linewidth). We have examined the underlying microstructural issues related to these observations. Thirdly, we point out that a combination of these techniques enables the ability to engineer thin film properties such as PL linewidth and spectral diffusion. Carefully engineering the structures, we have been able to demonstrate spectral diffusion of <200 MHz in 30 nm films on Si (within range of reports of 45 MHz and 110 MHz for Er in silicon[34]); and inhomogeneous linewidths down to ~5 GHz in anatase $TiO_2$ films (~ 10X larger that the best results for (rutile) bulk[12]). These materials studies indicate that the Er:$TiO_2$/Si is a promising platform for silicon compatible, chip-scale solid-state approaches for quantum coherent devices.

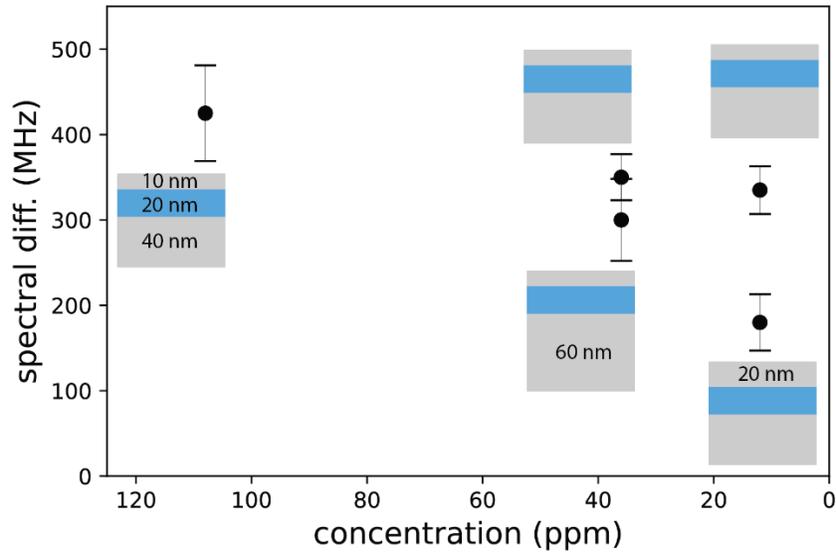

*Figure 6. The optical spectral diffusion bandwidth as a function of erbium concentration and the thickness of the top/bottom buffer layers. A layer structure accompanies each data point; unless otherwise noted, the top cap layer is 10 nm (gray); the doped region is 20nm (blue), and the bottom buffer is 40 nm (gray).*

## Acknowledgment

This work was primarily supported by the Center for Novel Pathways to Quantum Coherence in Materials, an Energy Frontier Research Center funded by the U.S. Department of Energy, Office of Science, Basic Energy Sciences under Award No. DE-AC02-05CH11231. M.K.S., S.G. and D.D.A were supported by this grant. Optical studies carried out by G.W. and S.E.S. were supported through the Quantum Metamaterials FWP by the U.S. Department of Energy; Office of Science; Basic Energy Sciences; Materials Sciences and Engineering Division. A.M. D. and J.H. acknowledge support by the U.S. Department of Energy, Office of Science, National Quantum Information Science Research Centers, under Award Number DE-FOA-0002253. Use of the Center for Nanoscale Materials, an Office of Science user facility, was supported by the U.S. Department of Energy, Office of Science, Office of Basic Energy Sciences, under Contract No. DE-AC02-06CH11357.